# Correlated Oxide Dirac Semimetal in the Extreme Quantum Limit

Jong Mok Ok[1,†,§], Narayan Mohanta[1,†], Jie Zhang[1,†], Sangmoon Yoon[1], Satoshi Okamoto[1], Eun Sang Choi[2], Hua Zhou[3], Megan Briggeman[4,5], Patrick Irvin[4,5], Andrew R. Lupini[1], Yun-Yi Pai[1], Elizabeth Skoropata[1], Changhee Sohn[1], Haoxiang Li[1], Hu Miao[1], Benjamin Lawrie[1], Woo Seok Choi[6], Gyula Eres[1], Jeremy Levy[4,5], Ho Nyung Lee[1*]

[1]Oak Ridge National Laboratory, Oak Ridge, TN 37831, U.S.A.

[2]National High Magnetic Field Laboratory, Florida State University, Tallahassee, FL 32310, USA

[3]Advanced Photon Source, Argonne National Laboratory, Lemont, IL 60439, USA

[4]Department of Physics and Astronomy, University of Pittsburgh, Pittsburgh, PA 15260, USA.

[5]Pittsburgh Quantum Institute, Pittsburgh, PA 15260 USA

[6]Department of Physics, Sungkyunkwan University, Suwon 16419, Korea

*Correspondence to: hnlee@ornl.gov

†These authors contributed equally to this work.

§Present address: Department of Physics, Pusan National University, Busan 46241, Korea

**Quantum materials (QMs) with strong correlation and non-trivial topology are indispensable to next-generation information and computing technologies. Exploitation of topological band structure is an ideal starting point to realize correlated topological QMs. Herein, we report that strain-induced symmetry modification in correlated oxide $SrNbO_3$ thin films creates an emerging topological band structure. Dirac electrons in strained $SrNbO_3$ films reveal ultra-high mobility ($\mu_{max} \approx 100,000$ cm²/Vs), exceptionally small effective mass ($m^* \sim 0.04 m_e$), and non-zero Berry phase. More importantly, strained $SrNbO_3$ films reach the extreme quantum limit, exhibiting a sign of fractional occupation of Landau levels and giant mass enhancement. Our results suggest that symmetry-modified $SrNbO_3$ is a rare example of a correlated topological QM, in which strong correlation of Dirac electrons leads to the realization of fractional occupation of Landau levels.**

Gaining control over the properties of quantum correlation and topology in quantum materials (QMs) is a critical step in advancing the physics of QMs that can promote innovation in many technological areas (*1-3*), such as spintronics and quantum technologies. Therefore, the discovery of QMs in which both quantum correlation and topology are achieved is a subject of current interest. Quantum spin liquids (*4-6*), topological superconductors (*7,8*), quantum anomalous Hall materials (*9,10*), and fractional quantum Hall systems (*11-16*) are good examples of correlated topological QMs whose properties depend on both correlation and topology. Moreover, strongly correlated materials highlighting many-body interactions are good examples of material systems in which to study entanglement. Thus, creating and understanding entangled states in QMs are important steps toward quantum technologies. One way to achieve entangled states in solids is to reach the extreme quantum limit (XQL) (*12,15-17*) (i.e., all carriers occupying entirely the lowest Landau level) at which correlations between charge carriers are maximized (*18,19*). At the XQL, the strong correlation gives rise to fractional quasiparticles, and some of the fractional states are expected to obey nonabelian statistics (*20,21*). The latter element is one of the most basic for topological quantum computing.

Chalcogenide- and halogen-based materials have been at the center of the development of topological QMs (*1-3,22-24*) because of their topological band structure and large spin-orbit coupling (SOC). However, the weak correlation in these topological materials hinders the appearance of key emergent phenomena, such as magnetism and superconductivity. Thus, the co-design approach, i.e., adding correlation to these topological materials, is currently under active investigation (*22-24*). A conventional approach includes both doping with $3d$ magnetic elements with strong correlation, and creating heterostructures with magnetic and/or superconducting materials (*7,9,10,22-24*). Such an approach, however, is a formidable task presenting a number of technical challenges.

In this context, transition metal oxides (TMOs) provide a versatile platform, owing to their inherent strong electron correlation and SOC (*25,26*), as well as advancements in their thin film synthesis and device fabrication. However, challenges also remain in studying TMOs, as the discovery of topological band structures is limited to only a couple of (or few) materials (*27-30*). Among oxide materials, $4d$ TMOs could be promising candidates, as they offer a good balance between electron correlation and SOC. Recently, a theoretical study predicted a topological band structure in an orthorhombic phase with $a^-a^-c^+$ octahedral rotations in SrNbO$_3$ (*31*). Dirac points in SrNbO$_3$, however, were predicted to exist far from the Fermi level. Although the electronic structure of perovskite oxides can be modified through the strain control of octahedral rotations (*32,33*), the viability of creating a novel Dirac semimetal with strained SrNbO$_3$ by manipulating the octahedral rotations has not been explored.

To experimentally realize the Dirac semimetallic state theoretically predicted in the $4d$ perovskite SrNbO$_3$ (*31*), an accurate determination of the crystallographic symmetry and its deliberate control are important. The utilization of epitaxial strain is known to be an effective method to modify the symmetry near the interface in a heterostructure (*32,33*). Although the crystallographic details of the perovskite SrNbO$_3$ have not been widely studied because of the difficulty of synthesizing the stoichiometric perovskite phase, a recent study with a polycrystal reported the lattice parameters of $a = \sqrt{2}a_p = 5.6894$ Å, $b = \sqrt{2}a_p = 5.6944$ Å, and $c = 2a_p = 8.0684$ Å ($\alpha = \beta = \gamma = 90°$) with an orthorhombic structure (space group: Pnma) (*34*). With a pseudocubic approximation, the lattice parameter $a$ is 4.023 Å (*34,35*). In this work, we grew SrNbO$_3$ films on lattice-mismatched SrTiO$_3$ substrates ($a_s = 3.905$ Å; therefore, the lattice mismatch of $\varepsilon(\%) = -3.02\%$) to induce epitaxial strain. To investigate the role of strain and the associated evolution of crystallographic symmetry, we systematically varied the thicknesses of films ($d = 2.4$–$130$ nm).

Epitaxial SrNbO$_3$ thin films were grown on (001) SrTiO$_3$ substrates by pulsed laser epitaxy. Despite the large lattice mismatch, high-quality SrNbO$_3$ thin films were successfully prepared, as was confirmed by several different methods, including reflection high-energy electron diffraction (RHEED), atomic force microscopy (AFM), x-ray diffraction (XRD), and cross-section scanning transmission electron microscopy (STEM) (see Figs. S1 and S2 in Supplementary Materials). The films revealed sharp interfaces without any detectable interfacial intermixing. In addition, the films were fully strained up to $d_{c1} = 7$ nm and partially strained from $d_{c1}$ to $d_{c2} = 18$ nm. Importantly, we observed that the $NbO_6$ octahedra were collectively distorted and rotated by the compressive strain. The fully strained film exhibited pronounced $c^-$ rotation-induced half-order peaks at (3/2 1/2 L/2), whereas these were absent in the fully relaxed thin films (Fig. 1C) (also see Fig. S3 in Supplementary Materials). The relaxed films ($d > d_{c2}$) exhibited a bulk-like cubic structure with a $a^0a^0c^0$ symmetry without any oxygen octahedral rotations (Fig. 1A), whereas the fully strained films ($d < d_{c1}$) showed a tetragonal structure with $a^0a^0c^-$-type octahedral rotations (Fig. 1B). The octahedral distortions found in strained thin films significantly affected the electronic structure of this material. In the cubic $SrNbO_3$, there is no Dirac-like degeneracy (Fig. 1D); but the four-fold degeneracy remains intact at two high-symmetry points, P and N, in tetragonal $SrNbO_3$ with the $a^0a^0c^-$ symmetry. Non-symmorphic symmetry, i.e., screw-rotation along the $z$ direction, is known to protect the degeneracy at these two high-symmetrical points (*36*) (see Fig. S4 and S5 in Supplementary Materials).

To investigate how epitaxial strain affects the transport properties of $SrNbO_3$ thin films, we first measured the temperature-dependent resistivity ($T = 2 - 300$ K) at the different thicknesses and zero magnetic field (Fig. 2A). Our films ($d = 2.4 - 27.2$ nm) exhibit clear metallic behavior down to 2.4 nm in thickness. Interestingly, a gradual reduction in the overall resistivity is observed as thin film samples become partially strained at around the second critical thickness $d_{c2}$ (Fig. 2B). In addition, the resistivity drastically increases at the ultrathin limit below $d_{c1}$ (Figs. 2A and B). The change in resistivity is attributed to a significant reduction in the carrier density (Fig. 2C and Figs. S6 and S7 in Supplementary Materials), supporting a strain-induced electronic structure modification in $SrNbO_3$ ultrathin films, as expected from density functional theory (DFT) calculations. Note that such a huge change in carrier density is unusual; but it is reminiscent of topological materials, in which 2–3 order-of-magnitude changes in the carrier density have been reported when a transition from the topological phase to the correlated phase (e.g. superconductivity) occurred (*37,38*). On the other hand, the resistivity of the relaxed films was around ~80 $\mu\Omega\cdot$cm at room temperature, which is of the same order of magnitude as a $SrNbO_3$ film grown on a $KTaO_3$ substrate with a smaller lattice mismatch ($\varepsilon \sim -0.85\%$) (*34*). These results strongly suggest that the observed transport properties are intrinsic to $SrNbO_3$, ruling out a possible contribution from oxygen vacancies in $SrTiO_3$ substrates (see Fig. S2 in the Supplementary Materials).

Having established that transport behaviors were the intrinsic properties of $SrNbO_3$ thin films, we measured the Hall effect of our films with a wider range of film thickness ($d = 2.4 - 74$ nm) to further understand the strain effect on the electronic state. As shown in Fig. S6 in the Supplementary Materials, three different types of Hall effects are observed when the thickness is varied. Fully relaxed films ($d > d_{c2}$) show a linear Hall effect, from which we extracted the carrier density ($n = 1/|e|\cdot dB/d\rho_{xy}$) and mobility ($\mu=(|e|\rho_{xx}n)^{-1}$). The carrier densities of the relaxed films were $n \approx 1.29 \times 10^{22}$ cm$^{-3}$, consistent with both the calculated value from a $d^1$ electron configuration of $n^{\text{theory}} \approx 1.53 \times 10^{22}$ cm$^{-3}$, and the values of the $SrNbO_3/KTaO_3$ samples mentioned previously (*34*). The mobility of the relaxed films was $\mu \approx 10$ cm$^2$/V·s at 300 K, which was again in good agreement with the $SrNbO_3/KTaO_3$ samples (*34*). The low-temperature mobility of our samples, however, was 100 times higher than that of the $SrNbO_3/KTaO_3$ samples because of the huge dielectric screening of $SrTiO_3$ at low temperature (*39*). Unlike the relaxed films, the strained samples ($d < d_{c2}$) exhibited nonlinear Hall effects. The nonlinear Hall effects can be explained with a two-carrier model by introducing an additional carrier (*40*), which presumably arises from the Dirac dispersion. The thickness-dependent carrier density and mobility are summarized in Figs. 2C and D. Note that the highest mobility observed is $\mu^{max} \approx$ 100,000 cm$^2$/V·s. This mobility value is higher than that of a correlated Dirac semimetal CaIrO$_3$ single crystal (27) and SrTiO$_3$ single crystals (41). The mobility of SrNbO$_3$/SrTiO$_3$ is 10 times higher than that of bulk SrTiO$_3$-related materials ($\mu^{STO} \sim$ 10,000 cm$^2$/V·s) because of the effective mass difference ($m^{STO}/m^{SNO} \sim 10$), which is described below.

More interesting, the additional small carriers evolve from holes into electrons as a function of thickness (Fig. 2C). Below 7 nm, the additional electron carrier has an extremely small carrier density of $n \approx 1.5 \times 10^{18}$ cm$^{-3}$ and a high mobility of $\mu \approx 10,000$ cm$^2$/V·s. When the three-fold valley and spin degeneracy are considered, the small electron carrier corresponds to a Fermi surface size of $S_F \approx 3.15 \times 10^{-4}$ Å$^{-2}$, of which the quantum limit is $H^*_{QL} \sim 3.3$ T. Thus, these carriers under this XQL offer an opportunity to study the strongly interacting topological electrons and their entangled states. Also note that the magnetic field required to reach the quantum limit in strained SrNbO$_3$ is only ~3.3 T, which is immensely lower than that for any conventional conducting oxides ($H^*_{QL} \sim 10^5$ T) (42).

To further investigate the quantum limit of a strained SrNbO$_3$ thin film ($d = 6.4$ nm), we studied magnetotransport properties at 0.15–10 K up to 30 T. To check the reproducibility of our results, four different samples (named S1–S4) were investigated (see Fig. S8 in the Supplementary Materials). All samples show clear quantum oscillations with qualitatively similar behaviors. First, we rotated the direction of the applied magnetic field with respect to the [001] axis and measured the quantum oscillations at several different angles ($\theta = 0 - 90^\circ$) (Fig. 3A). The quantum oscillations become more apparent in their second derivatives ($-d^2\rho/dH^2$) (Fig. 3B). The quantum oscillations survive at higher angles and do not exhibit $1/\cos\theta$ scaling; this behavior is attributed to the three-dimensional (3D) characteristics of the Fermi surface. The 3D character of the Fermi surface provides independent evidence to support that the transport properties originated from SrNbO$_3$ itself rather than from the interface with SrTiO$_3$ (see the Supplementary Materials for detailed discussions). The first evidence to support the quantum limit of SrNbO$_3$ thin films is an unsaturated linear magnetoresistance (MR) (Fig. 3C). At lower magnetic fields, the MR shows conventional $H^2$ behavior, but it changes to a linear MR at $H^*_{QL} \sim$ 3.3 T or higher. There are various possible mechanisms for this behavior. However, we attribute the unsaturated linear MR in the strained SrNbO$_3$ to quantum-linear MR (43) because it has a low carrier density. Such unsaturated linear MR is a characteristic of Dirac/Weyl semimetals at the quantum limit, as has been reported for many Dirac/Weyl semimetals (44-46).

Another important piece of evidence of the quantum limit was found in quantum oscillations. Conventional quantum oscillations for higher Landau levels are periodic in inverse magnetic fields ($1/H$). However, the quantum oscillations recorded for strained SrNbO$_3$ films exhibit an intriguing aperiodic behavior (Figs. 4A and B). This anomaly in the quantum oscillations provides an important characteristic unique to strained SrNbO$_3$ films. We attribute this quantum transport anomaly to the fractional occupation of the Landau levels owing to the strong correlation. As illustrated in Fig. S9 in the Supplementary Materials, we found that the Landau fan diagram ($1/H$ versus $N$) deviates significantly from the linear dependence when it is plotted by assigning the minima of resistivity—i.e., the second derivative of $\rho_{xx}$ ($-d^2\rho/dH^2$)—to integer numbers ($N$). When the $\rho_{xy} >> \rho_{xx}$ in a SrNbO$_3$ thin film is considered, the minima of $\rho_{xx}$ can be assigned to $N$ because $\sigma_{xx}=\rho_{xx}/(\rho_{xx}^2+\rho_{xy}^2) \sim A\rho_{xx}$, where A is a pre-factor. Such a strong deviation is obviously different from conventional quantum oscillations, even considering Zeeman splitting. On the other hand, the Landau fan diagram exhibits a clear linear relationship when the minima of the quantum oscillations are assigned to the rational fractions of the Landau integers $N$ = 1/3, 2/3, 4/3, 5/3, 2/5, 3/5, 4/5, 3/7, 4/7, 5/7, 6/7, and 4/9 (Figs. 4A and B) (also see Fig. S9 in Supplementary Materials). To ensure the reproducibility of this intriguing phenomenon, we tested four different samples and it was observed persistently in all the samples (see Supplementary Materials Figs. S8 and S9, and Fig. 4C). We further note that while the origin is still under debate, similar aperiodic quantum oscillations were reported in the 3D topological semimetal ZrTe$_5$ (16,17).

That entire series of fractional numbers of Landau integers seen in our strained perovskite oxide is highly extraordinary, as only extremely clean systems, including graphene (11) and GaAs-based 2D electron gases (12), have revealed such a complete set of fractional states. Only a portion of the fractional numbers have been observed in previously reported 2D systems at the quantum limit, such as a silicon quantum well (13), MnZnO/ZnO (14), and the surface state of Bi$_2$Se$_3$ (15). We attribute the observation of the complete set of the factional Landau levels from the highly strained SrNbO$_3$ ultrathin films to their high mobility and strong correlation.

In addition, from the Landau fan diagram (Fig. 4C), we were able to determine the Fermi surface size SF and Berry phase $\varphi_B$ from the Lifshitz-Onsager quantization rule $S_F(\hbar/eH)=2\pi(n+1/2-\varphi_B/2\pi+\delta)$, where $\delta$=0 for 2D or $\delta=\pm1/8$ for 3D. Based on this rule, we estimated the Fermi surface size, $S_F \sim 3.3\pm0.2$ T $\approx (3.15\pm0.19)\times10^{-4}$ Å$^{-2}$, which was in good agreement with the Hall data. Moreover, the Landau fan diagram confirms a nonzero Berry phase, as was predicted from the DFT calculations. The intercept of the Landau fan diagram was $(\varphi_B/2\pi+\delta)$=0.26$\pm$0.01. Note that it is possible that the nonzero Berry phase originated from the surface Rashba bands of SrTiO$_3$, as recently observed in ultra-thin LaTiO$_3$/SrTiO$_3$ (3–4 u.c.) heterostructures (47,48). However, that is unlikely to be the case in this SrNbO$_3$/SrTiO$_3$, because SrNbO$_3$ is metallic, and the Fermi level lies well above the Rashba-type band crossing. Instead, the nonzero Berry phase observed in the SrNbO$_3$ thin film suggests that the 4$d$ electrons in SrNbO$_3$, orbiting in cyclotron motion while enclosing a Dirac point in $k$-space, interacted strongly with one another and consequently formed fractional occupation of Landau levels. Thus, the aperiodic quantum oscillations and non-trivial Berry phase observed are unique for strained SrNbO$_3$ at the XQL compared with other oxide materials reported so far (see Table S2 in Supplementary Materials).

The most interesting experimental finding arising from the strongly interacting electrons was temperature-dependent quantum oscillations. From the temperature-dependent quantum oscillations, the cyclotron mass $m^*$ can be extracted, based on the Lifshits-Kosevich formula $\Delta\rho_{xx} \propto (\alpha m^* T/H)/\sinh(\alpha m^* T/H)$, where $\alpha=2\pi^2 c k_B \approx 14.69$ T/K. The temperature-dependent quantum oscillations depended highly on the applied magnetic fields (see Fig. S10 in the Supplementary Materials). Although quantum oscillations at low magnetic fields can be observed clearly at up to ~7 K, those at high magnetic fields are suppressed quickly with increasing temperature. Figure 4D shows the magnetic-field dependence of $m^*$ for four different samples. At lower magnetic fields, below $H^*_{QL} \sim 3.3$ T, the effective mass is around 0.04$m_e$—which is extremely light—as expected from the linear dispersion owing to the Dirac band of SrNbO$_3$ ($m^*_{DFT}$ ~0.026$m_e$). Interestingly, we observed a sudden mass enhancement of up to ~1$m_e$ at $H^*_{QL}$. Beyond the quantum limit, $m^*$ continuously increased to up to ~10 $m_e$ at 30 T. Note that, although mass enhancement has been reported in GaAs (49), ZrTe (16), and ZrSiS (50), those systems showed only a miniscule increase in effective mass of ~200–300 %.

Overall, these findings indicate that symmetry-modified $SrNbO_3$ is a 3D correlated oxide Dirac semimetal entering the XQL, in which topology meets many-body physics, yielding fractional occupation of Landau levels. Thus, we think that $SrNbO_3$ offers a promising platform for further exploration of exotic correlated quantum phases and behaviors that can provide innovative materials solutions for the next generation of quantum technologies.

## Materials and Methods

### Film growth

Substrates of $TiO_2$-terminated $SrTiO_3$ (001) $5 \times 5$ mm$^2$ in size were prepared by etching with buffered hydrofluoric acid and annealing at 1000 °C for 1 hour. A ceramic target was prepared by sintering mixtures of stoichiometric amounts of $SrCO_3$ and $Nb_2O_5$ powder at 1,100 °C for 10 hours, with an intermediate grinding and pelletizing step after the initial decarbonation step at 1,000 °C for 12 hours. A KrF excimer laser ($\lambda = 248$ nm) was used to ablate the target at a repetition rate of 5 Hz. $SrNbO_3$ thin films were grown on $SrTiO_3$ (001) substrates at optimum conditions of $T_g = 650$ °C, $P_{O2} < 4 \times 10^{-6}$ Torr, and $f = 0.4$ J/cm$^{-2}$. Owing to the 5+ preference of the niobium valence, the perovskite phase ($Nb^{4+}$) is achieved only under precisely controlled conditions of oxygen pressure ($P_{O2} < 10^{-5}$ Torr) (*34,35*), similar to the conditions needed for $LaTiO_3$ (*51,52*).

Film thickness was calibrated by x-ray reflectivity using a four-circle diffractometer. Crystallinity and epitaxial strain were examined by x-ray diffraction (XRD) and reciprocal space mapping. The surface morphology measurements were made with an atomic force microscope (AFM) (Veeco Dimension 3100). Supplementary Materials Fig. S1A shows XRD patterns of the $SrNbO_3$ 002 peak at different thicknesses ($d = 2.4$–15.2 nm). We observed a clear peak position shift at the critical thickness of $d_{c1} \sim 7$ nm, which is consistent with the huge jump in low-temperature resistivity (see Figs. 2A and 2B). Our thin films had atomically flat surfaces, as confirmed by the AFM image and the reflection high-energy electron diffraction image shown in Supplementary Materials Figs. S1B and S1C, respectively.

### Charge accumulation at the interface from STEM–EELS analysis

Cross-sectional transmission electron microscope (TEM) specimens were prepared using low-energy ion milling at $LN_2$ temperature after mechanical polishing. High-angle annular dark field (HAADF) scanning TEM (STEM) measurements were performed on a Nion UltraSTEM200 operated at 200 kV. The microscope was equipped with a cold field emission gun and a corrector of third- and fifth-order aberrations for sub-Ångstrom resolution. A convergence half-angle of 30 mrad was used, and the collection inner and outer half-angles for HAADF STEM were 65 and 240 mrad, respectively. A collection aperture of 5 mm was used for electron energy loss spectroscopy (EELS) measurement, and EELS spectrum imaging was performed at a speed of 30 frames per second.

Supplementary Materials Fig. S2A shows a HAADF STEM image of a $SrNbO_3$($d = 6.5$ nm)/$SrTiO_3$ sample viewed along the [100] zone axis. The HAADF STEM image confirms that the $SrNbO_3$ thin film is epitaxially grown on the $SrTiO_3$ substrate with a sharp interface. Note that the nonuniform background contrast in the HAADF STEM image can be attributed to the amorphized surface of the TEM specimen, which was formed by argon ion milling. Supplementary Materials Figs. S2B and S2C show the integrated Ti-$L_{2,3}$ and O-K edge EELS spectra across the interface. As shown in Supplementary Materials Fig. S2B, the spectral features of the Ti-$L_{2,3}$ edge broaden with the approach from the substrate to the interface. Note that the spectral change is most prominent at the first unit cell of the substrate, and the EELS spectrum fully returns to the spectrum of standard $SrTiO_3$ from three u.c. below. This spectral change is attributed to the charge transfer from the $SrNbO_3$ thin film.

The broadening of the Ti-$L_{2,3}$ edge spectra in $SrTiO_3$ could be due to any of the following three reasons (53): symmetry change at the interface, charge reduction due to oxygen vacancies, or charge reduction due to charge transfer from hetero-materials. Based on the HAADF STEM image, the first possibility can be ruled out because a sharp interface with a robust epitaxial relationship is observed. To further examine the origin of the charge reduction, we estimated the ratio between the titanium and oxygen continuum parts from the substrate bulk to the interface (Supplementary Materials Fig. S2D). The continuum parts were employed for the stoichiometry analysis because the near-edge structure was sensitive to the internal electronic states. As shown in Supplementary Materials Fig. S2D, there was no significant difference between the interface and the substrate bulk. Rather, the ratio became higher at the interface layer of $SrTiO_3$, which can be attributed to the beam broadening (the signals from $SrNbO_3$ can additionally contribute to the O-K edge). The stoichiometry analysis suggested that the spectral change at the interface (1 u.c. of $SrTiO_3$) was mainly due to the charge transfer from the $SrNbO_3$ thin film.

In addition to the EELS results, there are several other pieces of evidence that the charge transfer from the $SrNbO_3$ thin film, not oxygen vacancies, induced the charge accumulation at the interface. During pulsed laser deposition, vacancies can be formed when high-energy plasma plumes collide with the substrate. Therefore, the formation of oxygen vacancies depends to a great extent on the growth conditions, mainly the oxygen pressure and laser fluence. Higher vacuum and laser fluence create more oxygen vacancies. Oxygen stoichiometric films, however, can be grown even in a high vacuum if the laser spot size and energy are optimized, as our group reported earlier (54). We used a laser with a low enough fluence ($f = 0.4$ J/cm$^{-2}$) to grow $SrNbO_3$ thin films. To check for a possible oxygen deficiency in the substrate, we grew $SrTiO_3$ film on a $SrTiO_3$ substrate in the same growth conditions used for the $SrNbO_3$ thin films. The $SrTiO_3$ thin films that were grown were totally insulating, supporting the hypothesis that interfacial charges are not related to oxygen deficiency. Thus, charge transfer was most likely the origin of interfacial electrons, as predicted by a theoretical study (55). Unlike oxygen vacancies, which can extend deep into the $SrTiO_3$ side (56), charge transfer is expected only at the interface (57) (maximum 2–3 u.c. of $SrTiO_3$), which matches our observations well.

Lattice symmetry and strain-induced octahedral rotation of $SrNbO_3$

As shown in Supplementary Materials Figs. S3A and S3B, we used x-ray Bragg rod diffraction L scans at four different H-K quadrants to check and determine the primary lattice symmetry of $SrNbO_3$ as a function of film thickness. We present the fully strained 7.2 nm and fully relaxed 130 nm films as representative results (we also similarly measured 1.8, 16, and 20 nm $SrNbO_3$ films). Both strained and relaxed $SrNbO_3$ films displayed four-fold rotational symmetry. The 7.2 nm film showed a tetragonal-like lattice symmetry ($c > a$) as a result of the compressive strain, whereas the 130 nm film revealed a cubic-like lattice symmetry ($c \approx a$).

Oxygen octahedral rotation (OOR)–induced perovskite lattice doubling produced a unique set of half-order superstructure Bragg peaks, which were used to determine the OOR pattern and quantify the rotation angles. For detailed OOR half-order-peak investigations with very weak signals, we carried out synchrotron XRD measurements at room temperature at beamline 33-ID-D at the Advanced Photon Source at Argonne National Laboratory. Monochromatized x-rays with a wavelength of 0.61992 Å were used, and a Pilatus 100 K photon-counting area detector was used to capture the weak half-order superstructure. To suppress the fluorescence signal of the $SrTiO_3$ substrate, higher-energy x-rays ($E$ = 20 keV, well above the Sr $K$ absorption edge) were chosen. Scattering geometric corrections and background subtractions using a photon-counting area detector were conducted for all films. We surveyed all possible types of OOR half-order peaks to determine the rotation pattern with Glazer notation. The total absence of (odd/2, even/2, odd/2), (odd/2, odd/2, even/2) and (odd/2, even/2, even/2) types of peaks ruled out the existence of in-phase (+) rotation along either the $a$ or $c$ lattice axis or the existence of perovskite A-site cation off-symmetry point displacement (not shown in Supplementary Materials Fig. S3). As shown in Supplementary Materials Fig. S3C, the (H/2 K/2 L/2) (H = K) type peaks are also absent for all $SrNbO_3$ films with different thicknesses and strain states, which suggests there is no $a^-$ or $b^-$ type in-plane, out-of-phase rotation for any of the $SrNbO_3$ films. As shown in Supplementary Materials Fig. S3D, the (H/2 K/2 L/2) (H ≠ K) type peaks can be observed only in strained $SrNbO_3$ films (including a partially strained 16 nm film).

We then attempted to quantify the (H/2 K/2 L/2) (H ≠ K) type half-order Bragg rod to estimate the rotation amplitude (γ) for each strained film, using the scattering structural factor calculation of the complete heterostructure with a confined overall scale factor (via the diffraction intensity at the (002) thin film peak). In Supplementary Materials Fig. S3E, the simulated OOR half-order Bragg rods with different rotation γ angles are compared with the measured data. Supplementary Materials Fig. S3F displays the thickness dependence of extracted $c^-$ octahedral rotation angle γ. More detailed information regarding the strain, symmetry, and octahedral rotation of $SrNbO_3$ thin films with different thicknesses is summarized in Supplementary Materials Table S1. Both γ angles for 7.2 nm and 16 nm strained films are close to 10°, which was used for the density functional theory (DFT) calculations. The ultrathin 1.8 nm film exhibits a relatively reduced γ angle, probably because of the proximity effect of octahedral connectivity imposed by the underlying $SrTiO_3$ substrate without any OOR ($a^0a^0a^0$).

<u>Band structure calculation</u>

First principles electronic structure calculations were carried out within the framework of DFT on a plane wave basis with Perdew-Burke-Ernzerhof (PBE) exchange correlation (*58*), as implemented in the QUANTUM ESPRESSO simulation code (version 6.5) (*59*). We used a 7×7×8 Monkhorst-Pack $k$-point mesh to discretize the first Brillouin zone and a plane wave cut-off of 600 eV, which were found to be sufficient to achieve convergence of the total energy. The energy convergence criterion was set to $10^{-6}$ eV during the minimization process of the self-consistent cycle. Starting with the experimental lattice constants $a$=5.518 Å, $b$=5.518 Å, and $c$=8.280 Å, full optimization the $SrNbO_3$ tetragonal crystal structure (space group I4/mcm [140]) was performed using the force convergence criterion of $10^{-3}$ eV/Å. The calculations were performed with spin-orbit coupling both turned on and turned off.

The calculated Fermi surfaces of $SrNbO_3$ in the cubic (relaxed) and the tetragonal (strained) phases are shown in Supplementary Materials Fig. S4. The three Fermi surfaces correspond to the three $t_{2g}$ orbitals of niobium. The calculated band structure and Fermi surfaces of the cubic $SrNbO_3$ are consistent with recent reports (*60,61*). The heavier electronic band constitutes a larger Fermi surface in both the cubic and the tetragonal phases. The Dirac point at the P point near the Fermi level in the tetragonal phase is visible in the Fermi surface (bottom left plot) in the form of a connection between two disjoint surfaces (small surface with blue top and bigger surface with yellow top) at the P point.

Since the Dirac point at the P point is closer to the Fermi level, it can act as a source of a non-trivial Berry phase in the presence of a magnetic field. On the other hand, the Dirac points at the N point appear at ~0.7 eV above the Fermi level. Although these Dirac points at the N point are energetically unfavorable, oxide heterostructures would offer an avenue for designing a Dirac metallic phase by tuning the Fermi level closer to such Dirac points by strain or chemical substitution.

The Fermi velocity and effective mass near the Dirac point were estimated to be $7.07 \times 10^7$ m/s and 0.026 $m_e$, respectively. The high Fermi velocity near the P point was also expected to give rise to a high carrier mobility in strained SrNbO$_3$ thin films.

Strain tunable Dirac metallic state in SrNbO$_3$

As observed in our lattice symmetry measurements and octahedral rotation-induced half-order superstructure diffraction measurements, the substrate strain in the thin-film limit stabilized SrNbO$_3$ into a tetragonal crystalline symmetry (space group: I4/mcm) in which the NbO$_6$ octahedra were rotated only in the $x$-$y$ plane along the $z$-axis. As discussed in the main text, Dirac points appeared at the P point and at the N point of the Brillouin zone in this tetragonal crystalline environment. Supplementary Materials Fig. S5 shows the band dispersions at three different levels of the octahedral rotation. The Dirac point at the P point is found to remain closer to the Fermi energy in all three cases. Interestingly, the three Dirac points at the N point come closer to the Fermi energy with increasing octahedral rotation, as shown in Supplementary Materials Fig. S5D-F, enabling them to be available in the electronic transport. The octahedral rotation, therefore, offers a route to tune the Dirac points in a controlled manner that can be efficiently engineered in oxide heterostructures by using the substrate strain.

Transport measurements

Electrical transport measurements in this work were conducted with three measurement systems: a 14 T Physical Property Measurement System (Quantum Design), an 18 T dilution refrigerator at the University of Pittsburgh, and a 30 T bitter magnet with a He$^3$ cryostat at the National High Magnetic Field Laboratory (Tallahassee, USA). Results from different systems and different samples were reproducible and consistent. Aluminum wire-bonded contacts with a Van de Pauw configuration were used for measuring magnetotransport properties.

Strain-tunable multi-band nature of SrNbO$_3$ thin films

Supplementary Materials Fig. S6A shows the normalized Hall resistivity $\rho_{xy}$ of SrNbO$_3$ thin films with different thicknesses. Although a fully relaxed thin film ($d$ = 74 nm) shows linear Hall effects, nonlinear Hall effects were observed in the strained films ($d < d_{c1}$). To explain the observed nonlinear Hall effects in the SrNbO$_3$ thin films, a semiclassical two band model was used:

$$\rho_{xy} = \frac{(\mu_e^2 n_h + \mu_h^2 n_e) + (\mu_e \mu_h B)^2 (n_e + n_h)}{e[\mu_e|n_e| + \mu_h|n_h|]^2 + [\mu_e \mu_h (n_e + n_h) B]^2} B,$$

with the restriction of zero field resistivity:

$$\rho_{xx}(0) = \frac{1}{e[\mu_e|n_e| + \mu_h|n_h|]},$$

where $n_e(n_h)$, $\mu_e(\mu_h)$ are the carrier density and mobility, respectively, for electron(hole) type charge carriers. Supplementary Materials Fig. S6B shows the magnetic field dependence of $\rho_{xy}(H)$ (black dots) with fitting curves (colored solid lines). The convex-shaped $\rho_{xy}(H)$ for a 6.4 nm thickness film was well explained by using two electron carriers, and the concave-shaped $\rho_{xy}(H)$ found in the 12.4 nm thickness film was captured by electron and hole carriers.

Supplementary Materials Fig. S7 shows the temperature dependence of the estimated mobility and carrier density for strained ($d = 6.4$ nm) and fully relaxed ($d = 74$ nm) SrNbO$_3$ thin films. The temperature dependence of the electron mobility for both samples shows typical behavior, which follows the Matthiessen rule: $\mu^{-1} = \mu_0^{-1} + \mu_{e-e}^{-1} + \mu_{LO}^{-1}$, where $\mu_0$ describes the temperature-independent scattering from impurities or interface roughness that dominates at low temperature; $\mu_{e-e} \propto T^2$ quantifies the electron-electron scattering for intermediate temperatures; and $\mu_{LO}$ is the highly temperature-dependent electron-phonon scattering term (62). The carrier densities of two different samples, however, show clearly different degrees of temperature dependence. Whereas the carrier density of the relaxed film does not change significantly, that of the strained film decreases by two orders of magnitude at low temperature. These observations further highlight the strain-induced electronic structure change. As explained earlier, two types of electrons contribute to the transport properties of the stained thin film: one with high mobility and low carrier density, and the other with low mobility and high carrier density. The former can be regarded as the origin of the quantum oscillation with the Dirac nature, as discussed in detail in the main text. The latter, however, may come from a simple parabolic band and may not be able to generate quantum oscillations because of a large scattering rate. Under such a multicarrier condition, the longitudinal and Hall conductivities can be estimated by the following equations:

$$\sigma_{xx} = \sum \frac{n_i e \mu_i}{1 + (\mu_i B)^2} \text{ and } \sigma_{xy} = \sum \frac{n_i e \mu_i^2 B}{1 + (\mu_i B)^2} \ .$$

Thus, the high-mobility electrons dominated the transport properties at low temperature, especially under magnetic fields. Thus, one of the carriers could reach the quantum limit even though strained SrNbO$_3$ has multiple carriers. The compounds ZrTe$_5$ and HfTe$_5$ are other good examples in which only one carrier among multiple carriers shows quantum limit behavior (63,64).

Aperiodic oscillations

Supplementary Materials Figs. S8A and S8B display the low-temperature magnetoresistance and Hall resistivity of different SrNbO$_3$ thin films with almost the same thickness, ~6.4 nm. All samples (S1–S4) show consistent oscillations and the convex-shaped Hall effect that can be explained by the presence of two electron carriers, as previously discussed. Note that the oscillations are seen only in the films that have convex-shaped Hall effects. This finding supports the idea that one of the electron carriers, with low carrier density and high mobility, contributes to the oscillations. The oscillations are more pronounced in the second derivative curves ($-d^2\rho/dH^2$), as shown in Supplementary Materials Fig. S8C. The minima of resistance are consistent with the minima of the second derivative, as displayed in Supplementary Materials Fig. S9A.

By assigning the minima in oscillations to an integer Landau level index ($N$), we plotted the Landau fan diagram as illustrated in Supplementary Materials Fig. S9C. The $1/H$ versus $N$ deviates significantly from a conventional linear dependence. Such unusual behavior cannot be explained even considering the strong Zeeman splitting with an enormous Lande *g*-factor. For the unconventional behavior to be explained as the effect of Zeeman splitting, at least the low–magnetic-field region would have to show linear behavior, and its extrapolated line should be passing through near $N$=0. However, we could not find any field region in which the Landau fan diagram showed liner behavior, even under a low magnetic field. Furthermore, the slope of the Landau fan diagram in a low–magnetic-field region is too steep to pass $N$=0. More importantly, the deviation from conventional linear dependence is observed even in the low–magnetic-field $H$ < 3 T, at which the Zeeman effects are negligible (Zeeman energy < 0.5 meV). The unusual periodicity of oscillations, however, can be understood if the fractional Landau levels are taken into account, as discussed in the main text.

Note that aperiodic oscillations were also reported in $LaAlO_3/SrTiO_3$ samples. They were attributed to apparent spin degeneracy (*65*) and magnetic-field–dependent electronic bands of titanium $d_{yz}/d_{xz}$ orbitals (*66*). Both cases were based on the large effective mass of 1.4–1.7 $m_e$ as experimentally confirmed. Unlike the behaviors of $LaAlO_3/SrTiO_3$ 2DEGs, the $SrNbO_3$ films had an extremely small effective mass and giant mass enhancement at high fields that cannot be explained by the scenarios used for $LaAlO_3/SrTiO_3$.

Strong mass enhancement at quantum limit

Supplementary Materials Fig. S10 illustrates the magnetic field dependence of oscillations at different temperatures, showing their clear temperature and field dependence. With the temperature dependence of oscillations, the effective mass $m^*$ can be estimated by using the Lifshits-Kosevich formula $\Delta\rho_{xx} \propto (\alpha m^* T/H)/\sinh(\alpha m^* T/H)$, where $\alpha=2\pi^2 ck_B \approx 14.69$ T/K. Supplementary Materials Fig. S10A show mass plots for oscillations under several magnetic fields. In conventional cases, the effective mass does not change as a function of magnetic field. However, the strong magnetic field dependence of the effective mass is clearly seen in strained $SrNbO_3$ films. The magnetic field dependence of the effective mass is summarized in Fig. 4.

Origin of transport properties

Although there was charge accumulation at the interface, the transport properties were dominated by the $SrNbO_3$ rather than by the interface. Supplementary Materials Fig. S11 shows the carrier density dependence of mobility for the $SrNbO_3/SrTiO_3$ thin films and $SrTiO_3$-related materials (*67*). $SrNbO_3$ itself is the metal, with a huge number of carriers, $n \sim 10^{22}$ cm$^{-3}$, which is 3–5 orders of magnitude larger than the number in the 2D electron gas (2DEG) in $SrTiO_3$-related materials. Furthermore, the carriers with a larger density, $n \sim 10^{21} - 10^{22}$ cm$^{-3}$, in $SrNbO_3$ show one order of magnitude higher mobility than the 2DEG. The difference results from the effective mass difference between the two materials ($m^* \sim 0.1$ $m_e$ for $SrNbO_3$, $m^* \sim 1–2$ $m_e$ for 2DEG in $SrTiO_3$). The transport properties observed in $SrNbO_3/SrTiO_3$, therefore, originated from the $SrNbO_3$ itself rather than from the interface, because of the larger carrier density and higher mobility of $SrNbO_3$.

In addition, the observed quantum oscillations were not related to the interfacial electrons for the following reasons. First, the effective mass estimated from quantum oscillation is too light to be explained with a 2DEG. The typical effective mass of a $SrTiO_3$-based 2DEG is ~1–2 $m_e$ as found in $LaAlO_3/SrTiO_3$ (*68*), $Al_2O_3/SrTiO_3$ (*69*), and so on. Second, the quantum oscillation captures the 3D feature of the Fermi surface. The EELS spectrum directly verified that the excess electrons were mostly localized at the first monolayer of the substrate, which cannot account for the 3D Fermi surface. The 3D character can arise if the electron wavelength is smaller than the thickness. Based on the de Broglie relation, the electron wavelength is $\lambda = h/(m^* \cdot v)$. For a typical 2DEG in $SrTiO_3$, the wavelength is estimated to be ~0.7 nm ($v = 10^6$ m/s and $m^* = 1\ m_e$ for typical electrons). Thus, a 1–2 unit cell of $SrTiO_3$ exhibits a 2D character ($\lambda \geq d$) as previously reported (*68*). On the other hand, metallic $SrNbO_3$ can have a 3D Fermi surface. By considering the Fermi velocity of $SrNbO_3$ near the Dirac point $v = 7.07 \times 10^7$ m/s, which is an order of magnitude higher than the typical electron velocity of $10^6$ m/s, one can estimate the wavelengths of $SrNbO_3$ electrons. The wavelength of $SrNbO_3$ at a low field with $m^*$ ~0.1 $m_e$ is expected to be 0.1 nm ($\lambda \ll$ thickness); thus, $SrNbO_3$ could give rise to a 3D feature.

**Acknowledgments:** The authors thank J. S. Kim, Y. Kim, and M. Brahlek for discussion. **Funding:** This work was supported by the US Department of Energy (DOE), Office of Science, Basic Energy Sciences, Materials Sciences and Engineering Division, and in part by the Computational Materials Sciences Program. The high-magnetic-field measurements were performed at the National High Magnetic Field Laboratory, which is supported by National Science Foundation cooperative agreement no. DMR-1644779 and the state of Florida. This research used resources of the Advanced Photon Source, a DOE Office of Science User Facility, operated for the DOE Office of Science by Argonne National Laboratory under Contract No. DE-AC02-06CH11357. Extraordinary facility operations were supported in part by the DOE Office of Science through the National Virtual Biotechnology Laboratory, a consortium of DOE national laboratories focused on the response to COVID-19, with funding provided by the Coronavirus CARES Act. W.S.C. was supported by Basic Science Research Programs through the National Research Foundation of Korea (NRF) (NRF-2019R1A2B5B02004546). J.L. acknowledges support from NSF (PHY-1913034) and Vannevar Bush Faculty Fellowship (N00014-15-1-2847). **Author contributions:** J.M.O. and H.N.L. conceived the project. J.M.O. and J.Z. grew films and conducted characterization with help from E.S., C.S., H.L., H.M., W.S.C., and G.E. N.M.; and S.O. performed DFT calculations. S.Y. and A.R.L conducted STEM experiments. H.Z. conducted synchrotron x-ray diffraction measurements. J.M.O. performed transport measurements with help from J.Z., E.S.C., Y.Y.P., B.J.L., M. B., P.I. and J.L.. J.M.O. and H.N.L. wrote the manuscript with input from all authors. **Competing interests:** The authors declare no competing interests. **Data and materials availability:** All data are available in the main text or the Supplementary Material.

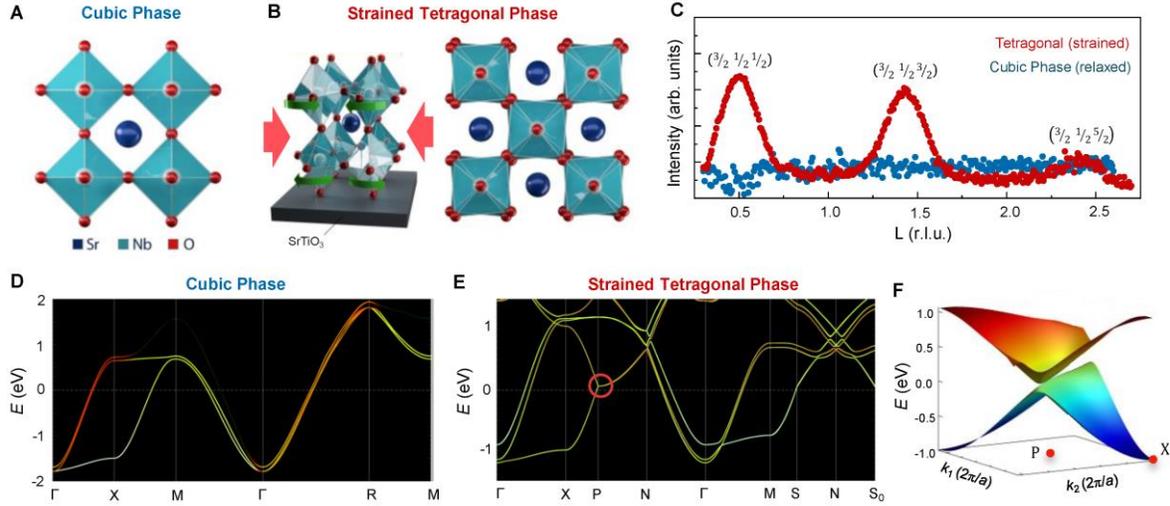

**Fig. 1. Strain-induced Dirac metallic state in SrNbO₃ thin films**. (**A-B**) Octahedral distortion pattern for (**A**) cubic $SrNbO_3$ ($a^0a^0a^0$ in the Glazer notation) and (**B**) strained tetragonal $SrNbO_3$ ($a^0a^0c^-$). Epitaxial strain induces octahedral distortion. (**C**) Octahedral rotation-induced half-order superstructure diffraction peaks of (3/2 1/2 L/2) with L = 1, 3, 5 for fully strained (red, 7.2 nm, $c^-$ rotation) and fully relaxed (blue, 130 nm, $c^0$ rotation) $SrNbO_3$ thin films. (**D**–**E**) Calculated electronic structure of (**D**) cubic $SrNbO_3$ and (**E**) strained tetragonal $SrNbO_3$. The red circle in plot (**E**) shows the Dirac point that appears near the Fermi level at the P point in the strained tetragonal phase. (**F**) Dirac dispersions near the P point within the tetragonal Brillouin zone. The larger Fermi velocity in the tetragonal phase, near the P point, would lead to higher carrier mobility and a favorable source of a non-trivial Berry phase in the presence of a magnetic field.

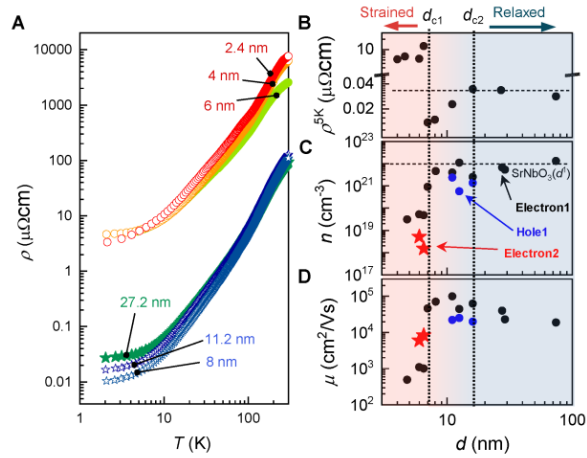

**Fig. 2. Thickness dependence of electronic state of SrNbO₃ thin film.** (**A**) Temperature dependence of resistivity of SrNbO₃ thin films with various thicknesses of 2.4 – 27.2 nm. (**B-D**) Thickness dependence of (**B**) resistivity, (**C**) carrier density and (**D**) mobility of SrNbO₃ thin films at 2 K. The carrier density of a relaxed thin film is well explained by the $d^1$ electron configuration. An additional electron carrier, which has high mobility $\mu \approx 10{,}000$ cm²/Vs, but extremely small carrier density $n \approx 1.5 \times 10^{18}$ cm⁻³, is observed in the strained thin films. The additional electron band has an experimentally reachable quantum limit of $H^*_{QL} \sim 3.3$ T because of its extremely small carrier density.

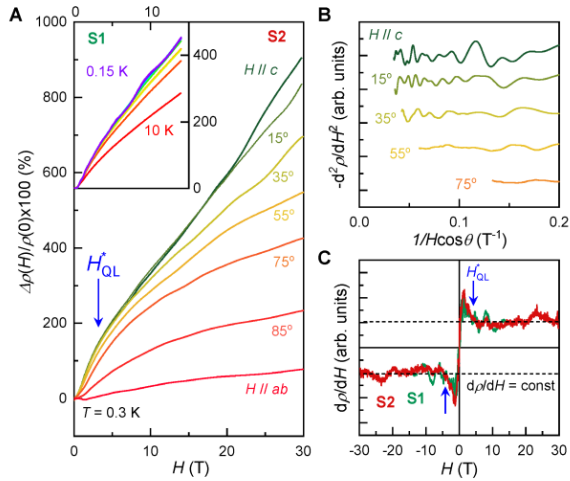

**Fig. 3. Angular dependence of quantum oscillations**. (**A**) Magnetoresistance for different angles up to 30 T at 0.3 K. Inset shows magnetoresistance for different temperatures ($T$ = 0.15 – 10 K). Quantum oscillations and linear magnetoresistance are clearly observed in SrNbO$_3$. (**B**) Second derivative of resistivity ($-d^2\rho/dH^2$) for different angles. Quantum oscillations are observed for all different angles and do not follow the $1/\cos\theta$ behavior, supporting the 3-dimensional character of the Fermi surface. (**C**) First derivative of resistivity ($d\rho/dH$) for two different samples at 0.3 K. The linear magnetoresistance starts to develop at $H^*_{QL}$ ~ 3.3 T.

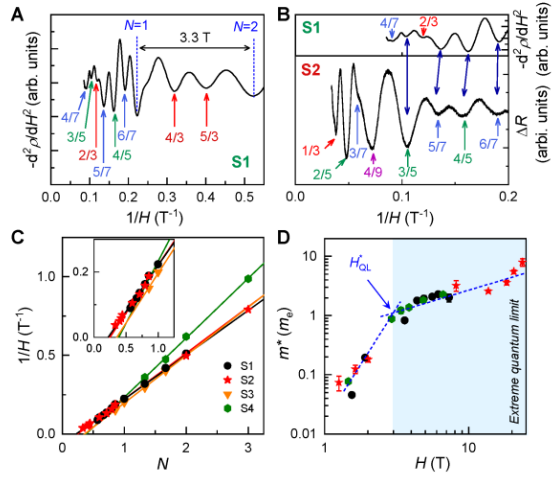

**Fig. 4.** Anomalous quantum oscillations in the quantum limit. (**A**) $-d^2\rho/dH^2$ as a function of $1/H$ under a magnetic field of up to 14 T for S1. The resistivity minima are assigned as integer (fractional) Landau levels, as indicated by the arrows. (**B**) $\Delta R$ as a function of $1/H$ under a magnetic field of up to 30 T for S2. S1 and S2 samples show consistent behavior. (**C**) Landau fan diagram of the Landau level index N versus $1/H$ for four different samples. All samples clearly show linear behavior. The inset shows an enlarged view of the high-field region. All samples have a non-trivial Berry phase as predicted by the calculations. (**D**) Effective mass at different magnetic fields for four different samples. Strong mass enhancement is found at $H^*_{QL} \sim 3.3$ T.

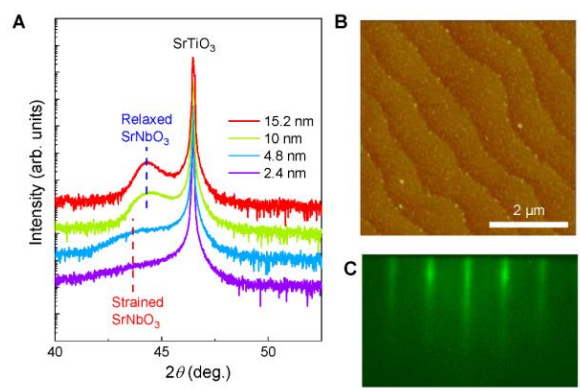

**Fig. S1.**

**Thin film growth of SrNbO₃.** (**A**) X-ray diffraction $2\theta$-$\theta$ patterns of SrNbO₃ thin films with different film thicknesses grown under optimum growth conditions. (**B**) Atomic force microscopy image of a SrNbO₃ thin film (6.4 nm in thickness). (**C**) Reflection high-energy electron diffraction image of a SrNbO₃ thin film (12 unit cells in thickness ~ 4.8 nm).

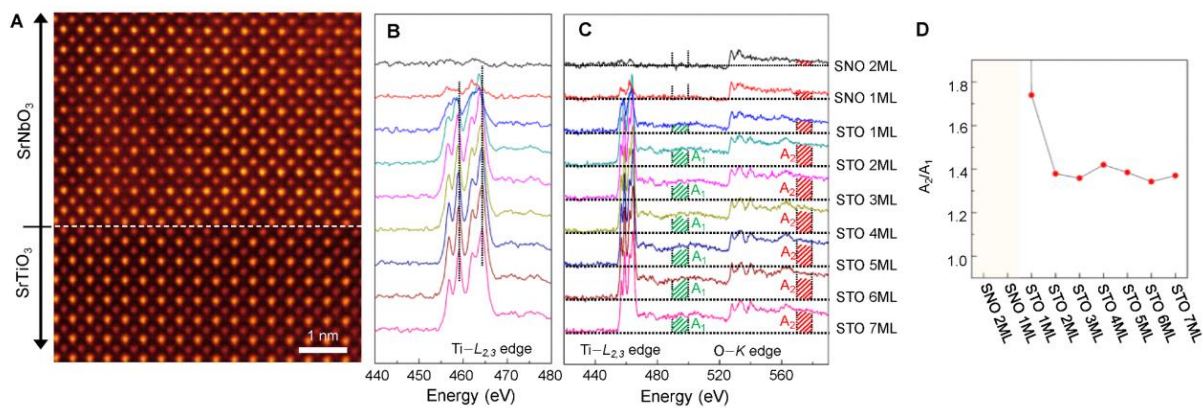

**Fig. S2.**

**Atomic and electronic structure of SrNbO₃/SrTiO₃ interface.** (**A**) High-angle annular dark field scanning transmission electron microscopy image of the SrNbO₃/SrTiO₃ interface viewed along the [100] zone axis. (**B**) Layer-resolved projected Ti-$L_{2,3}$ edge electron energy loss (EEL) spectra across the interface. (**C**) Layer-resolved projected Ti-$L_{2,3}$ and O-$K$ edge EEL spectra across the interface. The continuum part of the titanium and oxygen edges are denoted by A1 and A2, respectively. (**D**) The ratio of the Ti-$L_{2,3}$ and O-$K$ edge continuum across the interface.

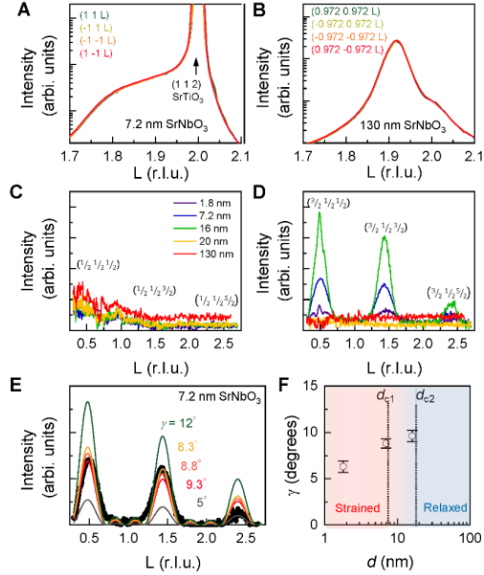

**Fig. S3.**

**Lattice symmetry and octahedral rotation-induced half-order super reflection of SrNbO₃ thin films.** (**A**-**B**) L scans at four different H-K quadrants through the 112 peaks for the (**A**) fully strained SrNbO₃ thin film (7.2 nm) and (**B**) fully relaxed SrNbO₃ thin film (130 nm), which indicate a four-fold rotational symmetry of the primary perovskite lattice. (**C**-**D**) Octahedral rotation-induced half-order peaks of (**C**) (1/2 1/2 L/2) type and (**D**) (3/2 1/2 L/2) type with L = 1, 3, 5 for SrNbO₃ thin films with different thicknesses. (**E**) Quantitative simulation curves of half-order Bragg rod of (3/2 1/2 L/2) for the 7.2 nm SrNbO₃ thin film. (**F**) Thickness dependence of extracted $c^-$ octahedral rotation angle γ.

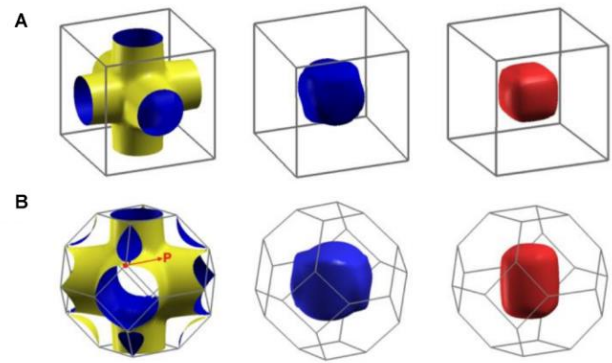

**Fig. S4.**
**Fermi surface of SrNbO₃.** (**A**) Top and (**B**) bottom rows show three Fermi surfaces of SrNbO₃ corresponding to (**A**) the cubic (relaxed) and (**B**) the tetragonal (strained) phases.

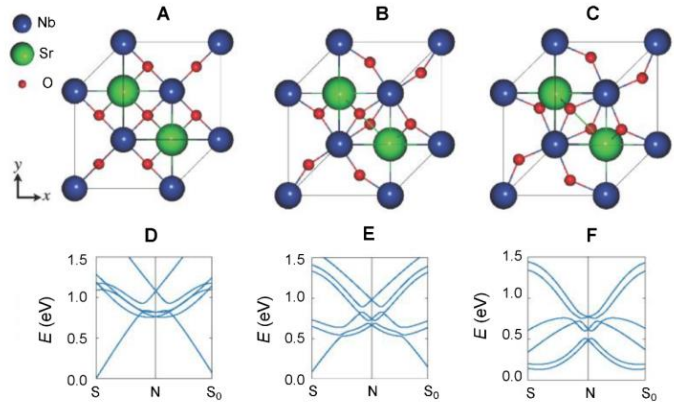

**Fig. S5.**

**Tunable Dirac dispersion by octahedral rotation.** The top row shows the top view of the tetragonal unit cell of $SrNbO_3$ in the I4/mcm crystalline symmetry at three different rotations of the $NbO_6$ octahedra: (**A**) zero rotation, (**B**) 10º rotation (the case discussed in the main text), and (**C**) 20 º rotation. The bottom row shows the corresponding band dispersions in **D**, **E**, and **F** along the S – N – S0 direction of the Brillouin zone. As the level of the octahedral rotation increases, the three Dirac points at the N point come closer to the Fermi energy.

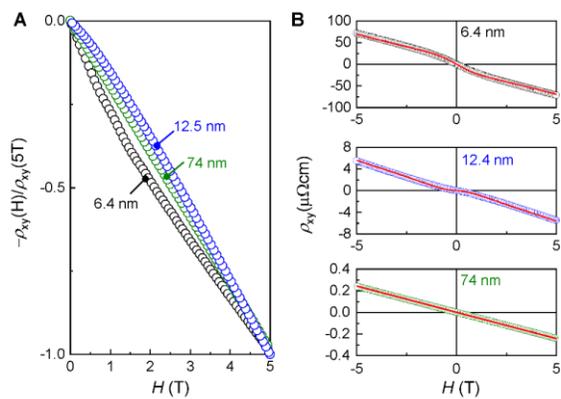

**Fig. S6.**
**Hall effect in a two-carrier model.** (**A**) Magnetic field dependence of normalized Hall effect at 2 K for samples of three different thicknesses. (**B**) Magnetic field dependence of Hall effects (dot) and fitted curves using two-carrier model (solid line).

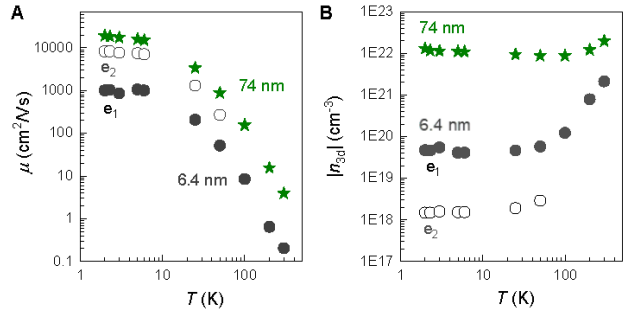

**Fig. S7.**
**Temperature dependence of mobility and carrier density.** Temperature dependence of (**A**) mobility $\mu$ and (**B**) carrier density $n_{3d}$ for strained ($d = 6.4$ nm) and relaxed ($d = 74$ nm) SrNbO$_3$ thin films. Two types of electron carriers found in the strained thin film are denoted as black open and closed dots, and an electron carrier for the relaxed film is marked by green stars.

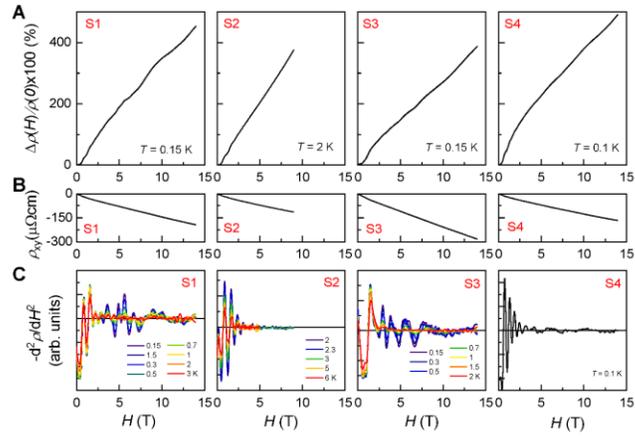

**Fig. S8.**

**Quantum transport of strained SrNbO₃ thin films.** (**A**) Magnetoresistance of strained SrNbO₃ thin films at low temperature (S1-S4). (**B**) Nonlinear Hall effect at low temperature, which can be explained with two electron carriers. (**C**) Magnetic field dependence of second derivative of resistivity at different temperatures.

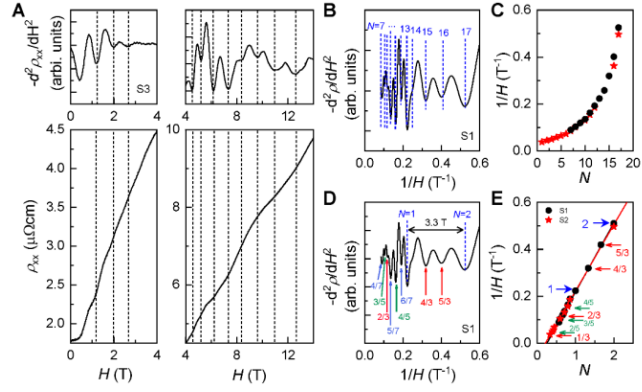

**Fig. S9.**

**Aperiodic oscillations of strained SrNbO₃.** (**A**) Magnetic field dependence of resistivity ($\rho_{xx}$) and its second derivative ($-d^2\rho/dH^2$). As marked by a dotted line, the minima of resistivity are consistent with the minima of the second derivative. (**B**–**E**) $-d^2\rho/dH^2$ as a function of $1/H$, in which the minima are assigned to the integer Landau level index (**B**) or fractional index (**D**). (**C**) and (**E**) show Landau fan diagrams corresponding to (**B**) and (**D**), respectively.

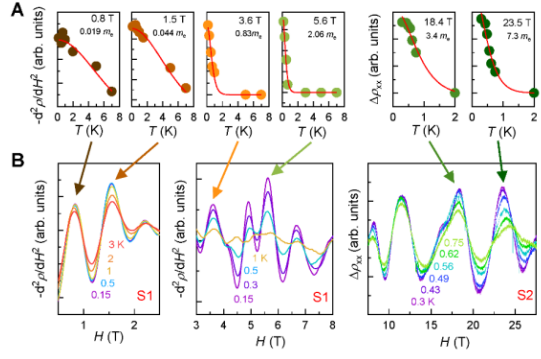

**Fig. S10.**

**Magnetic-field–dependent effective mass.** (**A**) Mass plot of several quantum oscillations under different magnetic fields. (**B**) Magnetic field dependence of $-d^2\rho/dH^2$ at different temperatures (0.15–3 K).

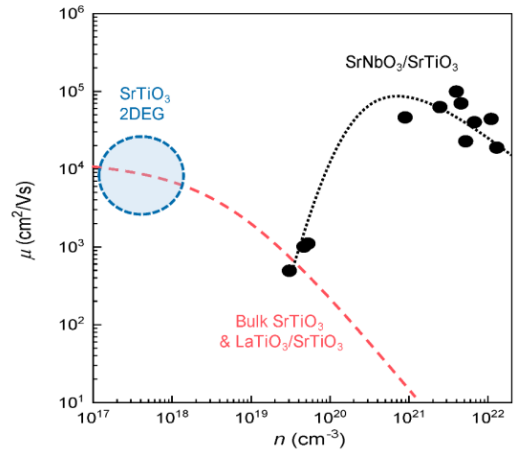

**Fig. S11.**

**Carrier density and mobility of SrNbO₃ thin films and SrTiO₃ related materials.** The carrier density dependence of the low-temperature electron mobility of SrNbO₃/SrTiO₃ and SrTiO₃ related materials (*67*).

**Table S1.**

**Strain, symmetry, and octahedral rotation of SrNbO₃ thin films**. Thickness, strained states, lattice symmetry, octahedral rotation pattern, and octahedral rotation angle of SrNbO₃ thin films.

| Thickness (nm) | Strained states | Lattice symmetry | Octahedral rotations | $\gamma$ angle (degrees) |
|---|---|---|---|---|
| 1.8 | Fully strained | Tetragonal-like | $a^0a^0c^-$ | $6.3 \pm 0.6$ |
| 7.2 | Fully strained | Tetragonal-like | $a^0a^0c^-$ | $8.8 \pm 0.5$ |
| 16 | Partially strained | Tetragonal-like | $a^0a^0c^-$ | $9.6 \pm 0.6$ |
| 20 | Partially relaxed | Tetragonal | $a^0a^0c^0$ | – |
| 130 | Fully relaxed | Cubic-like | $a^0a^0c^0$ | – |

**Table S2.**

**Quantum oscillations observed in oxide materials.** A summary of effective mass, Fermi-surface dimensionality, existence of Berry phase and aperiodicity, and origins of quantum oscillations of various oxide systems. Note that aperiodic oscillations found from 2DEGs of $LaAlO_3/SrTiO_3$ have different origins from those found in the $SrNbO_3$ films grown in this work, as explained in the Methods section.

| Origin of quantum oscillations | Material system | Effective mass ($m_e$) | Dimensionality | Berry phase | Periodic (P)/ aperiodic (AP) quantum oscillations |
|---|---|---|---|---|---|
| **Dirac semimetals** | **$SrNbO_3$ (This work)** | 0.04 | 3D | Non-trivial | AP |
| | **$CaIrO_3$(bulk)** (*27*) | 0.31 | 3D | Non-trivial | P |
| **Doped $SrTiO_3$** | **$SrTiO_3$(bulk)** (*70*) | 1.5, 6 | 3D | | P |
| | **$SrTiO_3$(bulk)** (*71*) | 1.82 | | | P |
| | **δ-doped $SrTiO_3$** (*72*) | 1.26 | 2D | | P |
| | **δ-doped $SrTiO_3$** (*73*) | 1.12–1.38 | 2D (27 nm), 3D (124 nm) | | P |
| | **$Sr(Ti,Zr)O_3/SrTiO_3$** (*74*) | 0.95–1.5 | | | P |
| **2DEG** | **$GdTiO_3/SrTiO_3$** (*75*) | 1 | | | P |
| | **$Al_2O_3/SrTiO_3$** (*69*) | 1.2 | | | P |
| | **$LaAlO_3/SrTiO_3$** (*68*) | 1.45 | 2D | | P |
| | **$LaAlO_3/SrTiO_3$** (*76*) | 0.92–2 | 2D | | P |
| | **$LaAlO_3/SrTiO_3$** (*65*) | 1.03–1.4 | 2D | | AP |
| | **$LaAlO_3/SrTiO_3$** (*66*) | 1.75 | | | AP |
| **Rashba spin splitting** | **$LaTiO_3/SrTiO_3$** (*47,77*) | 0.12, 1.2 | 3D | Non-trivial | P |
| | **$EuO/KTaO_3$** (*78*) | 0.59, 1 | 3D | Non-trivial | P |